\documentclass{elsart3}
\usepackage[square,comma]{natbib}
\usepackage{graphicx}
\usepackage{amssymb}
\usepackage{epsfig}
\usepackage{verbatim}
\usepackage{subfigure}
\usepackage{stmaryrd}

\newcommand{\bea}{\begin{eqnarray}}
\newcommand{\be}{\begin{equation}}
\newcommand{\ee}{\end{equation}}
\newcommand{\eea}{\end{eqnarray}}
\newcommand{\bfi}{\begin{figure}}
\newcommand{\efi}{\end{figure}}
\newcommand{\bc}{\begin{center}}
\newcommand{\ec}{\end{center}}
\newcommand{\nn}{\nonumber}

\begin{document}
\begin{frontmatter}

\title{Hidden Forces and Fluctuations from Moving Averages: \\A Test Study}

\author[fis,fis3]{V. Alfi}
\corauth[cor]{Corresponding author}
\ead{valentina.alfi@roma1.infn.it},
\author[fermi]{F. Coccetti},
\author[afs]{M. Marotta},
\author[fis,cnr,afs]{L. Pietronero},
\author[japan]{M. Takayasu}
\address[fis]{Universit\`a ``La Sapienza'', Dip. di Fisica, 00185, Rome, Italy.}
 % P.le A. Moro 5, 00185, Rome, Italy}
\address[fis3]{Universit\`a ``Roma Tre'', Dip. di Fisica, 00146, Rome, Italy.}
%  V. della Vasca Navale 84, 00146, Rome, Italy}
\address[cnr]{Istituto dei Sistemi Complessi - CNR, Via Fosso del Cavaliere 100, 00133 Rome, Italy.}
\address[fermi]{Museo Storico della Fisica e Centro Studi e Ricerche ``Enrico Fermi``, Via Panisperna, Rome, Italy.}
\address[afs]{Applied Financial Science, New York, USA.}
\address[japan]{Department of Computational Intelligence and Systems Science, Tokyo Institute of Technology,\\ Mail Box G3-52 4259 Nagatuta-cho, Yokohama 226-8502, Japan.}

%  \\ Interdisciplinary Graduate School of Science  and Engineering, \\Tokyo Institute of Technology, Mail Box G3-52 4259 Nagatuta-cho,\\ Yokohama 226-8502, Japan.}
%}
%/binampersand

\begin{abstract}
The possibility that price dynamics is affected by its distance
from a moving average has been recently introduced as  new statistical tool.
The purpose is to identify the tendency of the price dynamics to be attractive
or repulsive with respect to its own moving average. We consider a number of
tests for various models which clarify the advantages and
limitations of this new approach. The analysis leads to the identification of
an effective  potential with respect to the moving average. Its specific
implementation requires a detailed consideration of various effects which can 
alter the statistical methods used. However, the study of various model 
systems shows that this approach is indeed suitable
to detect hidden forces
in the market which go beyond usual correlations and volatility clustering.
\end{abstract}


\end{frontmatter}

\section{Introduction}

The concept of moving average is very popular in empirical trading
algorithms~\cite{bib1} but, up to now, it  has received little attention from a scientific
point of view~\cite{bib2,bib3,bib4}. Recently we have proposed that a  new definition of roughness
can be introduced by considering fluctuations from moving averages
with different time scales~\cite{bib5}.
This new definition seems to have various advantages with respect
to the  usual Hurst exponent 
in describing the fluctuations of high frequencies stock-prices. 

\begin{figure}[h]
  \begin{center}
    \resizebox{65mm}{!}{\includegraphics{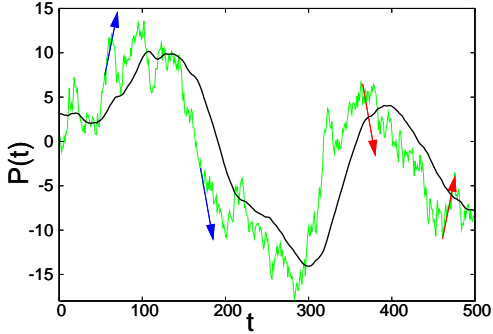}} 
    \caption{Example of a model of price dynamics (in this case a simple 
      random walk) together with its moving average defined as the average
      over the previous $50$ points. The idea is that the distance of the
price from its moving average can lead to repulsive (blue arrows) or
      attractive (red arrows) effective forces.}
    \label{fig1}
  \end{center}
\end{figure}

A more specific analysis of these fluctuations can be found in two recent
papers~\cite{bib6,bib7} which attempt to determine the tendency of the price to be
attracted or repelled from its own moving average (Fig. \ref{fig1}). 
This is completely different from the use of moving averages in finance, in
which empirical rules and predictions are defined in terms of a priori
concepts~\cite{bib1}. The idea is instead to introduce a statistical framework which is
able to extract these tendencies from the price dynamics.

\section{The Effective Potential Model}
The basic idea is to describe
price dynamics  in terms of an active random walk (RW) which is influenced by
its own moving average. This induces complex long range correlations
which cannot be determined by the usual correlation functions
and that can be explored by this new approach~\cite{bib6,bib7}.
The basic ansatz is that price dynamics $P(t)$ can be described in terms of a
stochastic equation of type:
\bea
\nn
&P&(t\!+\!\!1\!)\!-\!P(t)\!=\\
\nn
&=&\!-b(t)\frac{d}{d\Big(P(t)\!-\!P_M(t)\Big)}\Phi \Big(P(t)\!-\!P_M(t)\Big)\!+\\
&+&\sigma(t)\omega(t)
\label{eq1}
\eea
where $\omega(t)$ corresponds to a random noise with unitary variance
and
\bea P_M(t)\equiv\frac1M \sum_{k=1}^{M}P(t-k)
\label{eq2}
\eea
is the moving average over the previous $M$ steps.

The potential $\Phi$ together with the pre-factor $b(t)$ describe
the interaction between the price and the moving average. In both
approaches~\cite{bib6,bib7}
it is  assumed to be quadratic:
\bea
\phi\Big(P(t)-P_M(t)\Big)={\Big(P(t)-P_M(t)\Big)}^2.
\label{eq3}
\eea
Despite this similar starting point the two studies  proceed
along rather different perspectives. In Ref.~\cite{bib6} 
the three essential parameters of the model $(b; M; \sigma)$ are 
considered as constants with respect to $t$.
Then, by analyzing the price fluctuations over a suitable 
time interval and for a long time series, the values of the
three parameters are identified.

In Ref.~\cite{bib7}
instead the analysis is performed by looking directly
at the relation between $P(t+1)-P(t)$ and $P(t)-P_M(t)$.
This permits to derive the form of the potential and
to identify the
parameter $b(t)$ and its time variation.
For the US\$/Yen exchange rates the potential is found to be quadratic and it
is possible to rescale it with the term $1/(M-1)$ observing a good
data collapse.
This would imply that it is not necessary to specify the time scale
of the moving average.

\section{Test Studies}

Given these  different perspectives, which arise 
from the same basic model, we decided to perform 
a series of tests of this approach which we present 
in this paper.
We believe that these tests can elucidate various properties and
limitations of the new approach and represent a useful
information for its future developments and applications.

\begin{figure}[h]
  \begin{center}
    \resizebox{65mm}{!}{\includegraphics{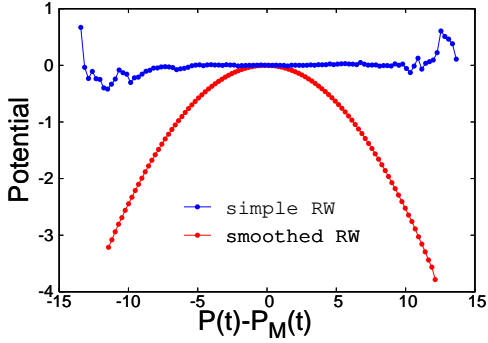}} 
    \caption{Effective potential for a random walk (flat line) and a smoothed
      random walk (convex parabola). The apparent repulsive potential
      corresponding to the smoothed RW is spurious and due
to the correlations corresponding to the smoothing procedure. The units of the
      potential are defined by Eq.(\ref{eq1}).}
    \label{fig2}
  \end{center}
\end{figure}

In Fig. \ref{fig1}
we show a simple RW and a moving average which represents its own smoothed
profile. The analysis is performed by plotting the values of
$P(t+1)-P(t)$ as a function of $P(t)-P_M(t)$ and deriving the potential
by integrating from the center~\cite{bib7}. 
The simple RW leads to a flat potential (no force) as  expected (Fig. \ref{fig2}). 
Than we can take the smoothed profile (previous moving average) as a dataset
by itself and repeat the analysis by comparing it to a new, smoother moving
average (not shown). As one can see in Fig. \ref{fig2}
this leads to an apparent repulsive potential which should be considered
as spurious.
\begin{figure}[h]
  \begin{center}
    \resizebox{65mm}{!}{\includegraphics{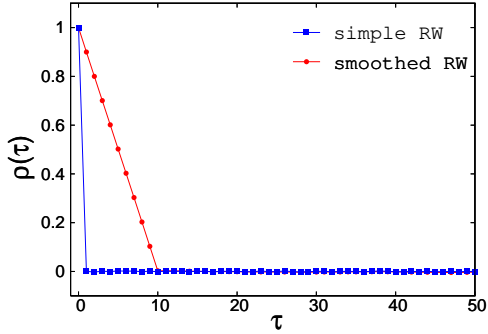}} 
    \caption{Autocorrelation ($\rho$) of the price increments for the simple RW
      and its smoothed profile. One can see that the smoothing procedure
      induces positive correlations up to the smoothing length (in this case
      $10$ steps).}
    \label{fig3}
  \end{center}
\end{figure}
This is due to the fact that the smoothed
curve implies some positive correlations as shown in Fig. \ref{fig3}. 
Therefore in this framework positive correlations lead to a destabilizing 
potential with respect to the moving average.
\begin{figure}[h]
  \begin{center}
    \resizebox{65mm}{!}{\includegraphics{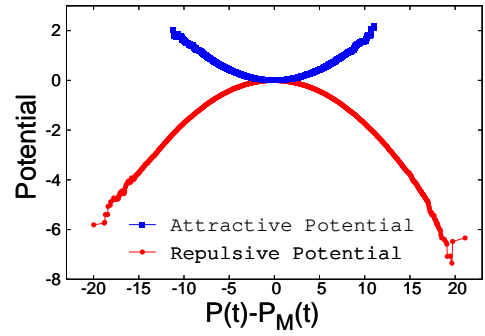}} 
    \caption{Effective potential reconstructed from a series of data obtained
      by
a dynamics corresponding to Eqs.(\ref{eq1}, \ref{eq2}) for the two cases of attractive
and repulsive potentials. In this case $M=30$ and $b=\pm 1$. The units of the
potential are defined by Eq.(\ref{eq1}).}
    \label{fig4}
  \end{center}
\end{figure}
The opposite would happen for negative correlations (zig-zag behavior).

The interesting question is however if one can identify
a non trivial situation in terms of the effective
potential but in absence of simple correlations.
 This would be the new, interesting situation and the corresponding
forces can be considered as hidden, in the sense that they do
not have any effect in the usual correlation functions.
Real stock-prices data clearly do not show any appreciable 
correlation, otherwise they would violate the simple arbitrage
hypothesis. In the exchange rates instead there is a zig-zag behavior
(negative correlation) at very short times which should be filtered with
suitable methods in order to perform the potential analysis~\cite{bib7}. 

\begin{figure}[h]
  \begin{center}
    \resizebox{65mm}{!}{\includegraphics{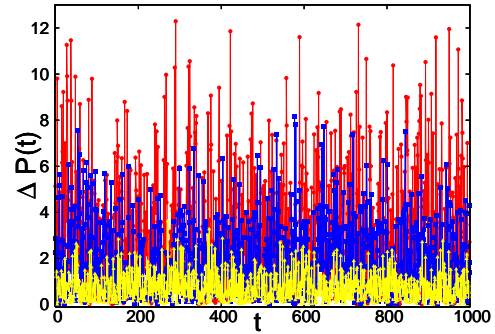}} 
    \caption{Absolute price variations for different time steps($\tau=1$ (yellow);
    $\tau=5$ (blue); $\tau=10$ (red)) corresponding to the dynamics of
    Eqs.(\ref{eq1}, \ref{eq2}).}
    \label{fig5}
  \end{center}
\end{figure}
\begin{figure}[h]
  \begin{center}
    \resizebox{65mm}{!}{\includegraphics{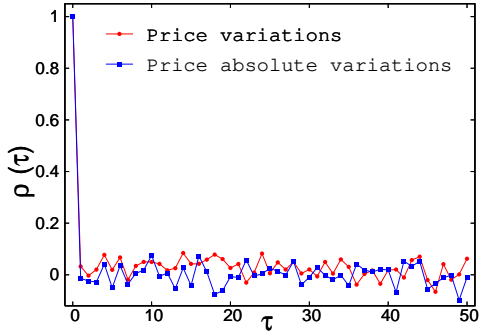}} 
    \caption{The correlation analysis of price variations shows no
      correlations between price differences and no volatility
clustering effect. 
This implies that the presence of  attractive or repulsive 
forces with respect to the moving averages is not detectable with the usual
statistical indicators.
}
    \label{fig6}
  \end{center}
\end{figure}

We now consider the model of the quadratic potential as in Refs.~\cite{bib6,bib7}.
The effective potential is easily reconstructed as shown in Fig. \ref{fig4}.
We also show in Fig. \ref{fig5}
the behavior of the absolute price variations for different time steps.
The correlation function for the price and volatility are shown 
in Fig. \ref{fig6}
which clarifies that, in this case, no simple correlation is present, nor
is there any volatility clustering effect. This is an interesting result
because it shows that the new method is able to detect hidden forces
which have no effect in the usual correlations of prices or volatility.

\section{Probabilistic Models}

\begin{figure}[h]
  \begin{center}
    \resizebox{65mm}{!}{\includegraphics{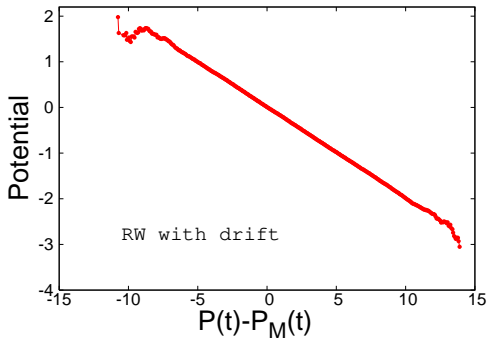}} 
    \caption{Effective potential corresponding to a RW with a constant drift
      which alters the probability for a step up or down. The units of the
      potential are defined by Eq.(\ref{eq1}).}
    \label{fig7}
  \end{center}
\end{figure}

We now consider some variations to the RW which depend on $P(t)-P_M(t)$.
We modify the probability of a certain step rather than
 the size of the step as in Eq.(\ref{eq1}).
The simplest model  is to add a constant drift, independent on the 
value of $P_M(t)$. The effective 
potential corresponding to this case is simply linear as shown in
Fig. \ref{fig7}. One can see that in this case the point 
where $P(t)-P_M(t)=0$ is not a special point and this model appears to be
oversimplified with respect to the dataset analyzed up to now~\cite{bib6,bib7}. 

\begin{figure}[h]
  \begin{center}
    \resizebox{65mm}{!}{\includegraphics{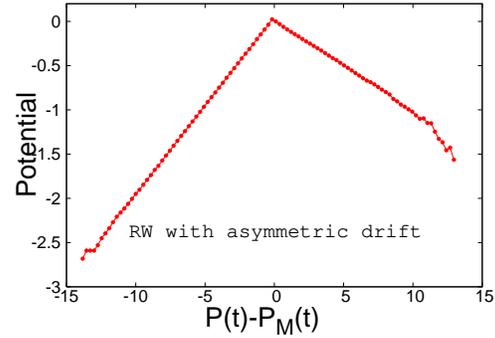}} 
    \caption{Effective potential corresponding to the dynamics of
      Eq.(\ref{eq4}) with $\epsilon_1=0.05$ and $\epsilon_2=0.10$. One can
      see that in this case the distribution is asymmetric and it extends more
      in the direction for which the instability is stronger. In this model
      the effective force only depends on the sign of $P(t)-P_M(t)$ and not on
      its specific value. The units of the
      potential are defined by Eq.(\ref{eq1}).}
    \label{fig8}
  \end{center}
\end{figure}

A more interesting model is represented by the following dynamics for 
a RW with only up and down steps:
\bea
\Bigg\{
\begin{array}{ll}
p(\uparrow)=1/2+\epsilon_1\;\;\;\mbox{for}\;\;\;P(t)-P_M(t)>0\\
p(\downarrow)=1/2-\epsilon_1
\label{eq4}
\end{array}\\
\Bigg\{
\begin{array}{ll}
p(\uparrow)=1/2-\epsilon_2\;\;\;\mbox{for}\;\;\;P(t)-P_M(t)<0\\
p(\downarrow)=1/2+\epsilon_2\;.
\nn
\end{array}
\eea
This implies a tendency of destabilization (repulsion from $P_M(t)$)
 whose strength is only dependent on the sign of $P(t)-P_M(t)$.
In principle the situation can be asymmetric with $\epsilon_1\neq\epsilon_2$.
The potential analysis for this case leads to a piecewise linear
potential in which the slopes are related to  $\epsilon_1$ and $\epsilon_2$ 
(Fig. \ref{fig8}).
\begin{figure}[h]
  \begin{center}
    \resizebox{65mm}{!}{\includegraphics{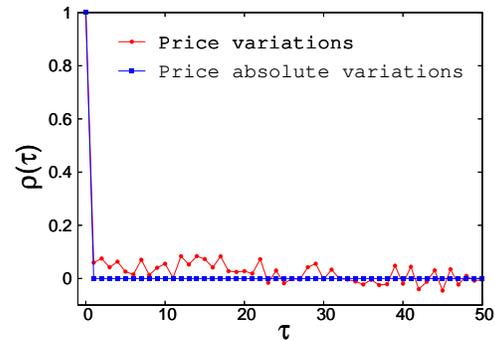}} 
    \caption{Correlation analysis of price variations and volatility
for the model of Eq. (\ref{eq4}). Also in this case no detectable correlations
are present.}
    \label{fig9}
  \end{center}
\end{figure}
One can also see that one line extends more than the
other indicating an asymmetric distribution. Also in this case 
the correlation of the price variations and volatilities show no detectable 
effect as shown in Fig. \ref{fig9}.
Clearly in this case the effective potential is just a representation of the
correlations between $P(t+1)-P(t)$ and  $P(t)-P_M(t)$ whose microscopic
origin is instead in the modification of the probability for unitary steps.

\section{Fractal Model}

\begin{figure}[h]
  \begin{center}
    \resizebox{65mm}{!}{\includegraphics{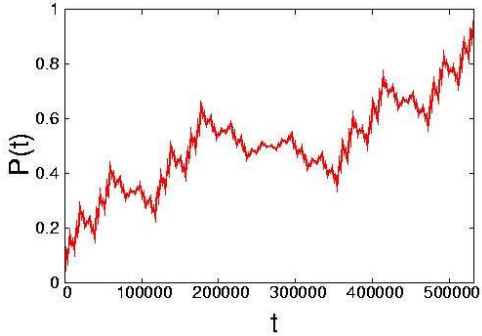}} 
    \caption{Example of a fractal model distribution of price}
    \label{fig12}
  \end{center}
\end{figure}

It may be interesting  to consider also  the case of a fractal model
constructed by an iterative procedure \cite{bib2}, Fig. \ref{fig12}.

\begin{figure}[h]
  \begin{center}
    \resizebox{65mm}{!}{\includegraphics{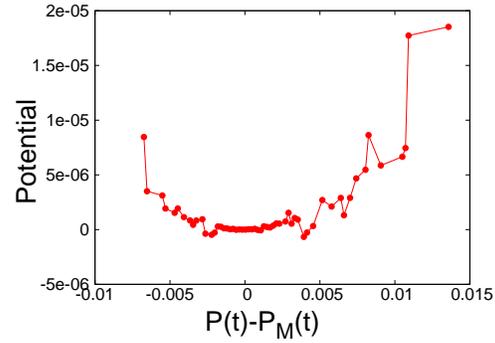}} 
    \caption{Effective potential corresponding to the fractal price model. The units of the
      potential are defined by Eq.(\ref{eq1}).}
    \label{fig13}
  \end{center}
\end{figure}

The fractal model does not have a specific dynamics but, since it is
often considered as to capture some properties of real prices, we 
consider of some interest to study if this model  would correspond to some
type of effective potential.
In Fig. \ref{fig13} we can see that the effective potential
is slightly attractive.
Given the symmetry of the model
construction, the asymmetry observed in the effective potential
is probably due to the backward construction of the
corresponding moving average.

\section{Analysis of the Fluctuations}

\begin{figure}[h]
  \begin{center}
    \resizebox{65mm}{!}{\includegraphics{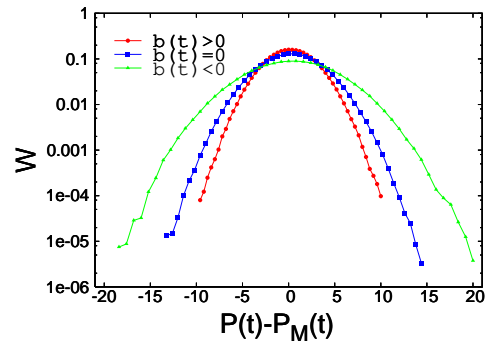}} 
    \caption{Distribution of the fluctuations, $W(P(t)-P_M(t))$, for the
      dynamics of Eq. (\ref{eq1}-\ref{eq3}) and different values of the
      parameter $b$.}
    \label{fig10}
  \end{center}
\end{figure}

We  now consider the nature of fluctuations from the moving average
by analyzing the probability distribution $W\Big(P(t)-P_M(t)\Big)$
for the various models. In Fig. \ref{fig10}
we show the distributions corresponding to the 
quadratic potential as compared to that of a reference RW $(b=0)$. The 
first observation is that the repulsive potential makes the 
distribution broader (super diffusion) while the attractive
potential makes it narrower (sub diffusion).
This behavior was already observed in Refs.~\cite{bib6,bib7}.
Less trivial is the fact that the distributions are well
represented by  gaussian curves. 
\begin{figure}[h]
  \begin{center}
    \resizebox{65mm}{!}{\includegraphics{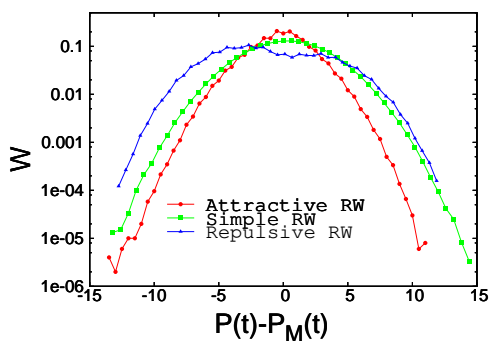}} 
    \caption{Distributions of the fluctuations, $W(P(t)-P_M(t))$, for the
      dynamics of Eq. (\ref{eq14}). In this case the distributions became
      asymmetric due to different values of ${\epsilon}_1$ and ${\epsilon}_2$.}
    \label{fig11}
  \end{center}
\end{figure}

In Fig. \ref{fig11}
we show the same distributions corresponding to the probabilistic model 
of Eq. (\ref{eq4})
for the case of asymmetric attractive and repulsive effects.
In this case there is a marked deviation from the gaussian behavior
and the case of repulsive trend develops two separate peaks.
It will be interesting to check the corresponding distribution
on real stock-prices which we intend to perform in the
future.

\section{Conclusions and Perspectives}

In summary the idea to consider price dynamics as influenced by an 
effective force dependent on the distance of price $P(t)$
from its own moving average $P_M(t)$ represents a new
statistical tool to detect hidden forces in
the market.
The implementation of the analysis can be seriously affected by
the eventual presence of positive or negative correlations.
However, we have shown by a series of models and tests, that
this new method is able to explore complex correlations which
have no effect on the usual statistical tools like
the correlations of price variation and the 
volatility clustering.

The method provides an analysis of the sentiment
of the market: aggressive for the case 
of repulsive forces and conservative for attractive ones.
In this respect it may represent
a bridge between the financial technical analysis and the application of
statistical physics to this field.
In addition it may also be  useful to analyze the results of the different
strategies
and behaviors which arise in  agent based models.


\begin{thebibliography}{99}

\bibitem{bib1}
B.J.~Murphy, Technical Analysis of the Financial Markets, Prentice Hall Press, 1999.



\bibitem{bib2}
B.B.~Mandelbrot, Fractals and Scaling in Finance, Springer Verlag, New York, 1997.


\bibitem{bib3}
R.N.~Mantegna, H.E.~Stanley, An Introduction to Econophysics, Cambridge
University Press, Cambridge, 2000. 

\bibitem{bib4}
J.P.~Bouchaud, Theory of Financial Risk and Derivative Pricing, Cambridge
University Press, Cambridge, 2003.


\bibitem{bib5}
V.~Alfi, F.~Coccetti, M.~Marotta, A.Petri, L.Pietronero, Roughness and Finite
Size Effect in the NYSE Stock-Price Fluctuations, preprint 2006.


\bibitem{bib6}
VR.~Baviera, M.~Pasquini, J.~Raboanary, M.Serva, Moving Averages and Price
Dynamics, {\em International Journal of Theoretical and Applied Finance},
vol.5, num. 6, pag. 575-583, 2002.


\bibitem{bib7}
M.~Takayasu, T.~ Mizuno, H.~Takayasu, Potentials of Unbalanced Complex
Kinetics Observed in Market Time Series, http://arxiv.org/abs/physics/0509020.




%%%%%%%%%%%%%%%%%%%
\end{thebibliography}
\end{document}